\documentclass[twocolumn]{aastex631}

\shorttitle{Formation of moons and equatorial ridge around top-shaped asteroids after surface landslide}
\shortauthors{Hyodo et al.}
\graphicspath{{./}{figures/}}

\begin{document}

\title{Formation of moons and equatorial ridge around top-shaped asteroids after surface landslide}


\author[0000-0003-4590-0988]{Ryuki Hyodo}
\affiliation{ISAS/JAXA, Sagamihara, Kanagawa, Japan}
\email{ryuki.h0525@gmail.com}

\author[0000-0002-3839-1815]{Keisuke Sugiura}
\affiliation{Earth-Life Science Institute, Tokyo Institute of Technology, Meguro-ku, Tokyo 152-8550, Japan}

\begin{abstract}
Top-shaped asteroids have been observed among near-Earth asteroids. About half of them are reported to have moons (on the order of $\sim 1$wt.\% of the top-shaped primary) and many of them have an equatorial ridge. A recent study has shown that the enigmatic top-shaped figure of asteroids (e.g., Ryugu, Bennu, and Didymos) could result from an axisymmetric landslide of the primary during a fast spin-up near the breakup rotation period. Such a landslide would inevitably form a particulate disk around an asteroid with a short timescale ($\sim 3$ hours). However, the long-term full dynamical evolution is not investigated. Here, we perform a continuous simulation ($\sim 700$ hours) that investigates the sequence of events from the surface landslide that forms a top-shaped asteroid and a particulate disk to disk evolution. We show that the disk quickly spreads and produces moons (within $\sim 300$ hours). The mass of the formed moon is consistent with what is observed around the top-shaped asteroids. We also demonstrate that an equatorial ridge is naturally formed because a fraction of the disk particles re-accretes selectively onto the equatorial region of the primary. We envision that Ryugu and Bennu could once have an ancient moon that was later lost due to a successive moon's orbital evolution. Alternatively, at the top-shaped asteroid that has a moon, such as Didymos, no significant orbital evolution of the moon has occurred that would result in its loss. Our study would also be qualitatively applicable to any rubble-pile asteroids near the breakup rotation period.
\end{abstract}


\section{Introduction}\label{sec_intro}

\begin{figure*}[ht]
\begin{center}
\includegraphics[width= \linewidth]{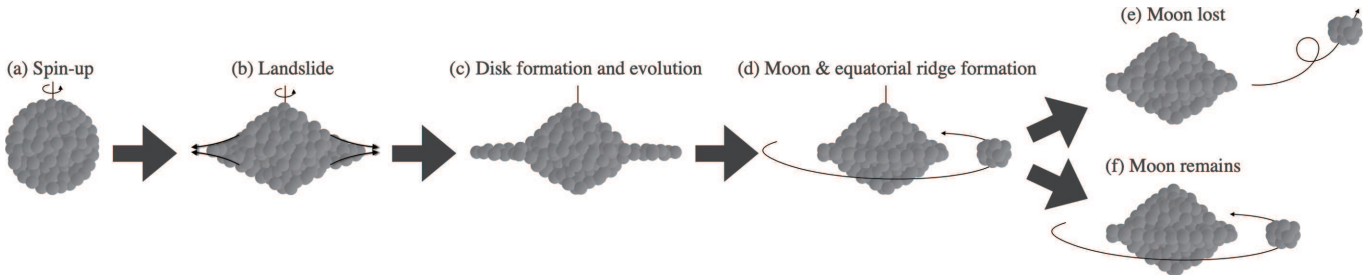} 
\caption{Schematic summary of our paper. Panel (a): a rubble-pile asteroid spins up due to, for example, the YORP effect, small impacts, a close encounter with a planet, or re-accumulation after a catastrophic impact. Panel (b): a surface landslide occurs when a critical spin-state is realized and a top-shaped figure is formed \citep{Sug21}. Panel (c): a particulate disk spreads due to inelastic collisions and gravitational interactions among particles. Panel (d): a moon is gravitationally accreted outside the Roche limit of the central top-shaped body and an axisymmetric equatorial ridge is formed due to the re-accretion of disk particles. Panels (e) and (f): the formed moon is lost or remains, depending on the long-term orbital evolution between the moon and the primary. \label{fig_summary}}
\end{center}
\end{figure*}

More and more growing interests are on near-Earth asteroids (NEAs) together with recent activities of asteroid exploration missions. Top-shaped asteroids may exist ubiquitously among NEAs \citep{Wal15,Mar15}. These include asteroid 162173 Ryugu \citep{Wat19}, 101955 Bennu \citep{Lau19}, 65803 Didymos, 1999 KW4 \citep{Ost06}, and 2001 SN263 \citep{Bec15}. Their diameters are less than several kilometers, and recent in-situ observations by spacecrafts showed that they might be rubble-pile bodies \citep[e.g.,][]{Wat19}. Because of their small size, Yarkovsky-O'Keefe-Radzievskii-Paddack (YORP) effect \citep{Rub00}, small impacts \citep{Tak09}, and/or a close encounter with a planet \citep{Hyo16}, for example, could efficiently change their spin states. The acceleration rate of spin-up (or spin-down) depends on each physical process (e.g., roughly instantaneous by an impact and $\gtrsim 100$ Kyr by YORP effect on a kilometer-sized body). 

Interestingly, about half of the reported top-shaped asteroids have moons around them (e.g., 2001 NE263; \citealt{Bec15}, 1999 KW4; \citealt{Ost06}, 1994 CC; \citealt{Bro11}, Didymos; \citealt{Nai20}). The masses of the moons are roughly on the order of $\sim 1$wt.\% of the host top-shaped asteroids. Furthermore, many of the top-shaped asteroids are reported to have an equatorial ridge \citep{Ben15}. 

There are several proposed mechanisms to form a top-shaped asteroid. The top-shaped figure may be formed during re-accretion after catastrophic disruption of a parent body \citep{Mic20}. Alternatively, a mass movement with reshaping or a landslide of surface materials of a rubble-pile asteroid due to fast spin-up may form the top-shaped figure \citep{Wal08,Wal12,Har09,Hir20,Sug21}.

Recently, \cite{Sug21} used the Smoothed Particle Hydrodynamics (SPH) method to study rotational deformation during spin-up, which has also been used to model the shape deformation due to impacts \citep{Jut15b}. In their study, they included effective bulk friction of rubble pile bodies as a parameter (i.e., effective friction angle $\phi_{\rm fri}$), which could effectively include, for example, the effect of cohesion (see Sec.~2.2 in \citealt{Sug21}). They demonstrated that, for $\phi_{\rm fri} \geq 70$ deg with a spin-up timescale of $\lesssim$ a few days (defined as the elapsed time of spin-up from the rotation period of 3.5 h to 3.0 h), an axisymmetric set of surface landslides would occur and top-shaped rubble-pile bodies would be correspondingly formed (panels (a)-(b) in Fig.~\ref{fig_summary}). 

The story of the landslide hypothesis, however, should not have ended here, although \cite{Sug21} did not study the fate of the surface materials of the landslide. In this study, we first show that the surface materials are distributed around the newly formed top-shaped central body, forming a transient particle disk (panel (c) in Fig.~\ref{fig_summary}). Here, the disk mass is typically found to be about $\sim 10$wt.\% of the top-shaped primary body. 

In this study, following the numerical approach used in \cite{Sug21}, we investigate a full long-term dynamical evolution -- from the landslide to the end of the disk evolution -- to understand the fate of the surface materials that are distributed around the newly formed top-shaped body (Fig.~\ref{fig_summary}). We show that moons and an axisymmetric equatorial ridge are naturally and inevitably formed around the top-shaped primary (panel (d) in Fig.~\ref{fig_summary}).

In Sect.~\ref{sec_method}, we describe our numerical models. In Sect.~\ref{sec_results}, we show our numerical results. Section \ref{sec_discussion} discusses the long-term orbital evolution between a moon and the central primary due to the binary YORP (BYORP) effect, which may lead to diverse systems of the top-shaped asteroids with and without a companion moon(s) (panel (e) and panel (f) in Fig.~\ref{fig_summary}). Finally, Section \ref{sec_summary} summarizes this paper.\\

\section{Numerical method}\label{sec_method}

Our SPH simulations are the same as those in \cite{Sug21}, and we refer the readers to their Sec.~2 for more details. Our simulations solve hydrodynamic equations (the equations of continuity and motion) with self-gravity and yield processes with the aid of a friction model that determines the shear strength of granular material \citep{Jut15a}. 

In a similar context, a Discrete Element Methods (DEMs) with cohesionless particles were used in \cite{Wal08}. Shape deformation using this approach could be numerically affected due to the inconsistency between the realistic particle size and that used in their simulations. In the SPH approach, in contrast, a rubble-pile body was constructed by a continuum of granular material and we explicitly set a specific value of the friction angle of the material. We expect that the SPH method has the advantage of precise control of bulk friction of rubble piles \cite[see also][]{Jut15b}. We also note that there is a difference in the process of the moon formation between their study and ours; in \cite{Wal08}, the moon formation occurred via a step-by-step accumulation of discretely ejected particles, while, in our study, the moon formation occurs because of particle disk evolution.

The rubble-pile body initially had a uniform and spherical body with a radius of $R_{\rm ini}=500$ m, which was then numerically spun up with the angular acceleration of $8.954 \times 10^{-10}$ rad s$^{-2}$. We stopped the acceleration when 1wt.\% of the primary is ejected; this prevents further artificial deformation after the landslide, and the arbitrary choice of 1wt.\% does not affect our results as long as the landslide occurs with a shorter timescale (here $\sim 3$ h) than the spin-up timescale (here $\sim 30$ h). The density of particles at uncompressed states is $\rho_{\rm ini}=1.19$ g cm$^{-3}$. Particle mass is given as $m=M_{\rm tot}/N_{\rm tot}$, where $M_{\rm tot} \equiv (4/3)\pi \rho_{\rm ini} R_{\rm ini}^3$ and $N_{\rm tot}=25470$ is the total number of the SPH particles.

In this study, following \cite{Sug21}, we specifically focused on a set of parameters (i.e., $\phi_{\rm fri}=80$ deg; see panels (k)-(o) in their Fig.~1), which successfully resulted in the formation of an axisymmetric top-shaped figure as seen at asteroids Ryugu, Bennu, and Didymos. Lowering $\phi_{\rm fri}$, for example, resulted in an elongated lemon-shaped primary via internal deformation (see panels (a)-(e) and (f)-(j) in Fig.~1 of \citealt{Sug21}), which eventually led to distribute particles as disks with a variety of initial disk mass. The dependences on the disk parameters (e.g., initial disk mass) and tidal parameters are studied in detail using an independent numerical approach (1D fluid simulations) in our companion paper (Madeira et al. submitted). Here, we used the direct and continuous approach -- from the landslide to the end of the disk evolution -- and focused on studying the details regarding the relevant consequences of the top-shape formation.  

Below, we show that the surface materials of the primary are distributed in a disk-like structure and such a particulate disk eventually forms moon(s) around the top-shaped primary. We numerically distinguished member particles of the primary as follows. We start from a randomly chosen particle and then iteratively detect any particles within a critical distance in a bottom-up fashion \citep{Hyo18}. This procedure continues until no more new particles are detected. As a critical distance, we used 1.5 times the smoothing length of an SPH particle to be consistent with \cite{Sug21}, although this specific choice does not affect our results. 

In this study, we continued our simulation up to $t \sim 25 \times 10^5$ s ($\sim 700$ hours). This is much after the landslide had occurred; \cite{Sug21} stopped their simulations at $t \sim 1 \times 10^5$ s ($\sim 30$ hours) when the top-shaped figure was formed. We may need to be careful about the angular momentum (AM) conservation in SPH simulations, particularly for a rotating system (such as disk evolution). In our calculations, after an artificial spin-up of the primary (i.e., artificial AM increase in the system) was stopped, the cumulative AM error was $\sim 2.6$\% during and after the disk formation until the end of the simulation (i.e., from $t \sim 1.5 \times 10^{5}$ s to $t \sim 25 \times 10^{5}$ s).

Although our SPH simulations contain some error in AM conservation, independent numerical simulations of particle disk evolutions of the Earth's Moon formation (e.g., $N$-body simulations; e.g., \citealt{Kok00} and 1D semi-analytical fluid simulations; e.g., \citealt{Sal12}) used similar initial disk conditions (e.g., in terms of disk to primary mass ratio) and their results (e.g., resultant moon to primary mass ratio) were consistent with those obtained here. This is because the gravitational disk evolution can be well characterized by the mass ratio between the disk and the primary \citep{Kok00,Sal12}. Furthermore, using $N$-body simulations, \cite{Hyo15} showed a scaling law for the mass of the largest moon formed via disk spreading as a function of the initial disk mass (see their Fig.~12) and it is generally consistent with that obtained in this study. We note that, although these studies generally assumed particulate disks, a more realistic Moon-forming disk would be vapor-rich and thus the Earth's Moon formation process would be more complex \citep[e.g.,][]{Tho88,Nak14,Loc18} \\

\begin{figure*}[t]
\begin{center}
\includegraphics[width=0.898\linewidth]{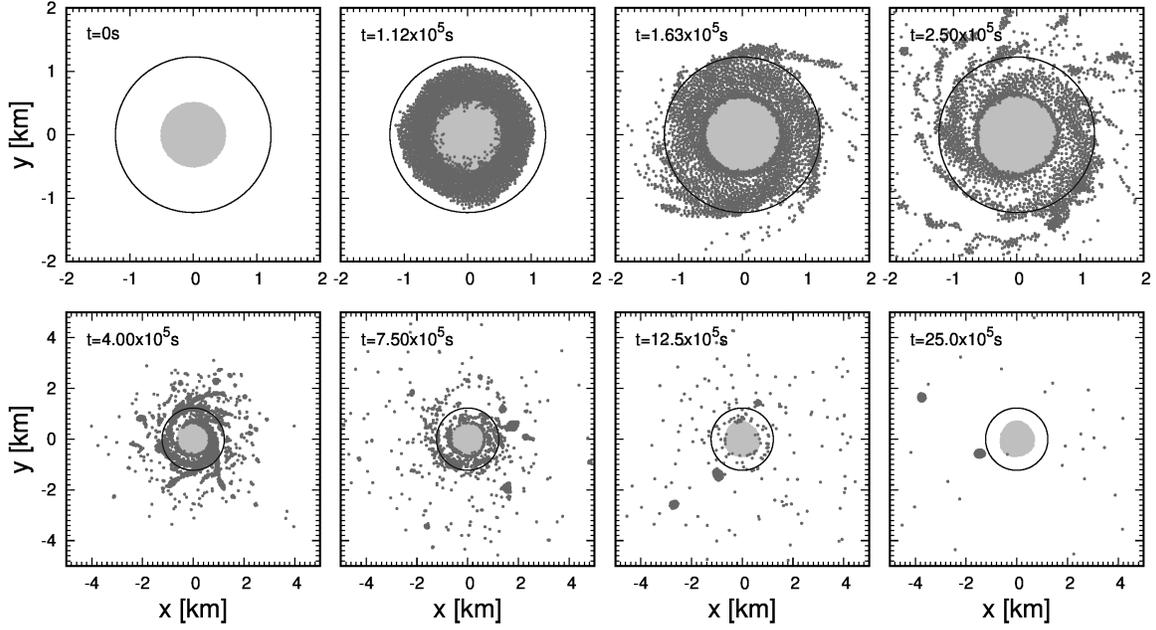} 
\caption{Overall disk evolution after the landslide. Gray particles represent member particles of the top-shaped primary.  Dark-gray particles represent disk particles and member particles of gravitational clumps (including the formed moon at later epochs). The black circle indicates the Roche limit of the primary ($r_{\rm R} \sim 2.5 R_{\rm c}$ where $R_{\rm c}=500$m is used). \label{fig_evo_xy}}
\end{center}
\end{figure*}

\begin{figure*}[htb]
\begin{center}
\includegraphics[width=0.898\linewidth]{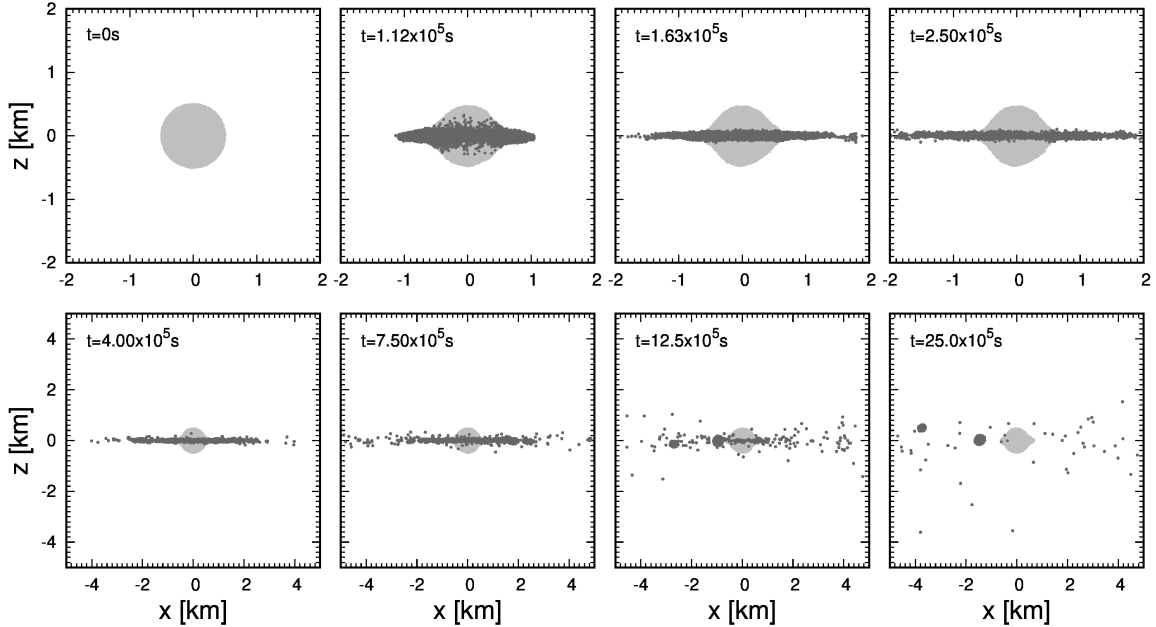} 
\caption{Same as Fig.~\ref{fig_evo_xy}, but edge on views. \label{fig_evo_xz}}
\end{center}
\end{figure*}

\section{Numerical results: moon formation}\label{sec_results}

\begin{figure*}[ht]
\plotone{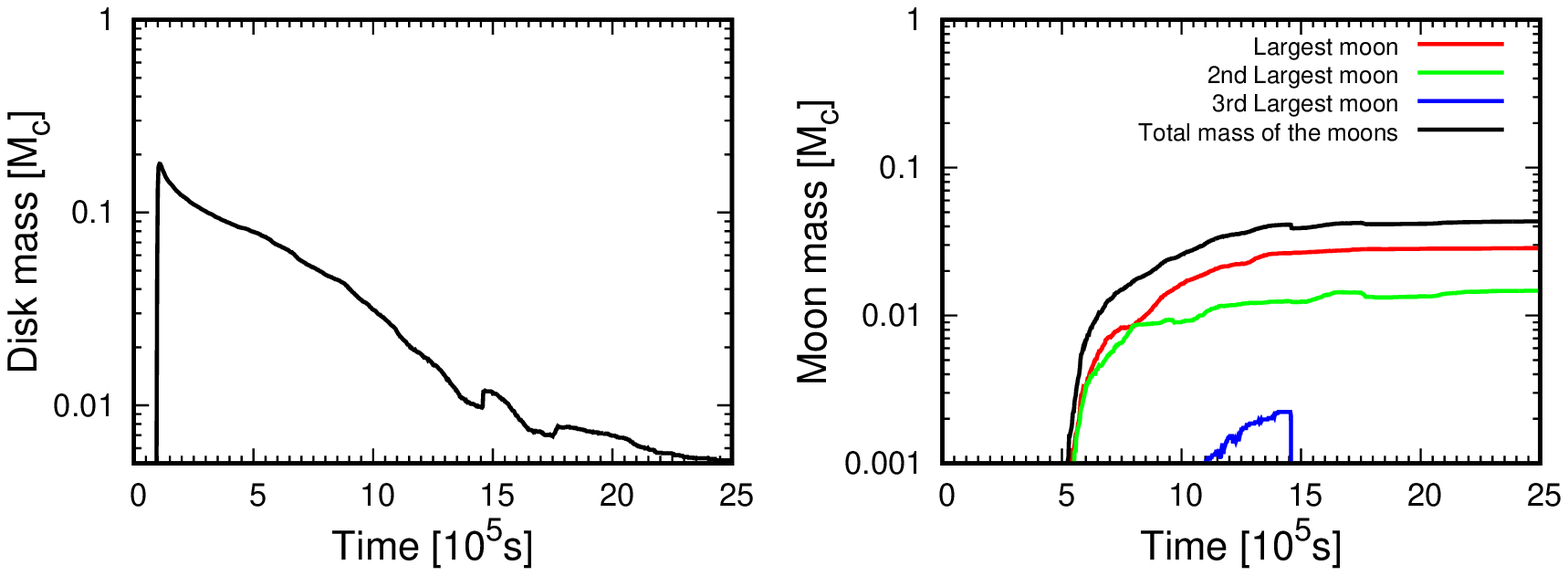}
\caption{Evolutions of the disk mass (left panel; not including the mass of the largest moons) and the masses of the largest three moons (right panel). At around $t \sim 1 \times 10^5$s with a time-interval of $\Delta t \sim 10^4$s, the landslide occurs and a particle disk is formed (left panel). The disk spreads and starts to form gravitational clumps with masses larger than $\sim 0.001M_{\rm c}$ at $t \sim 5.5 \times 10^5$s, where $M_{\rm c}$ is the mass of the central primary. Two large moons are formed by the end of our simulations ($t=25 \times 10^5$s; $\sim 700$h). \label{fig_evo_mass}}
\end{figure*}

Figures \ref{fig_evo_xy} and \ref{fig_evo_xz} show an overall evolution of a particle disk including the landslide. The landslide (i.e., disk formation) occurs with a very short timescale ($\sim 10^4$ s $\simeq 3$ hours) compared to the disk evolution timescale ($\sim 10^6$ $\simeq 300$ hours; see Fig.~\ref{fig_evo_mass}). Because of the landslide, a top-shaped figure of the primary is formed (see more details in \citealt{Sug21} and their Fig.1). The surface materials of the primary are distributed around the primary, forming a particle disk. The initial mass of the disk is $M_{\rm disk} \sim 20$wt.\% of the top-shaped primary (Fig.~\ref{fig_evo_mass}). Most of the disk mass is initially within the Roche limit of the primary, defined as $r_{\rm R} \equiv 2.456 (\rho_{\rm m}/\rho_{\rm c})^{-1/3}R_{\rm c}$ where $\rho_{\rm m}$ and $\rho_{\rm c}$ are densities of the moon and the central body. $R_{\rm c}$ is the radius of the central body. 

The disk mass is large enough to become gravitationally unstable and the spiral arm structures are formed as a result of gravitational instability. This leads to an efficient angular momentum transfer and the outer disk spreads further radially outward \citep{Tak01}, distributing the disk materials beyond the Roche limit. Because of the angular momentum conservation, the inner part of the disk spreads inward, resulting in re-accretion onto the primary. 

The materials scattered beyond the Roche limit start to coagulate via their own gravity, forming gravitational clumps (Figs.~\ref{fig_evo_xy} and \ref{fig_evo_xz}). With time, accretion among clumps proceeds, forming larger gravitational bodies (see also the right panel of Fig.~\ref{fig_evo_mass} for the mass evolution of the largest objects). After $\sim 300$ hours since the landslide, large moons around the top-shaped primary asteroid are formed. The total mass of the formed moons is $\sim 4$wt.\% of the primary. 

Here, two large moons are formed (Figs.~\ref{fig_evo_xy}, \ref{fig_evo_xz}, and \ref{fig_evo_mass}). The number of the large moon is the result of the stochastic nature of the accretion processes that can be seen in the 3D simulations, especially when the disk is massive to the central body; the details of the stochastic nature were reported in independent $N$-body simulations \citep{Hyo15}. This means that, if one runs another simulation with the same parameters but with slightly different initial positions of particles, the final outcome could be the formation of a single moon. Thus, the number of large moons is not a decisive parameter that characterizes the general outcome. More importantly, instead, regardless of the number of moons, the total mass of the moons rarely changes and it characterizes the outcome of specific choices of the disk parameters (e.g., initial disk mass; see \citealt{Ida97,Kok00}).\\

\section{Discussion} \label{sec_discussion}
\subsection{Formation of equatorial ridge} \label{sec_ridge}

\begin{figure*}[ht]
\plotone{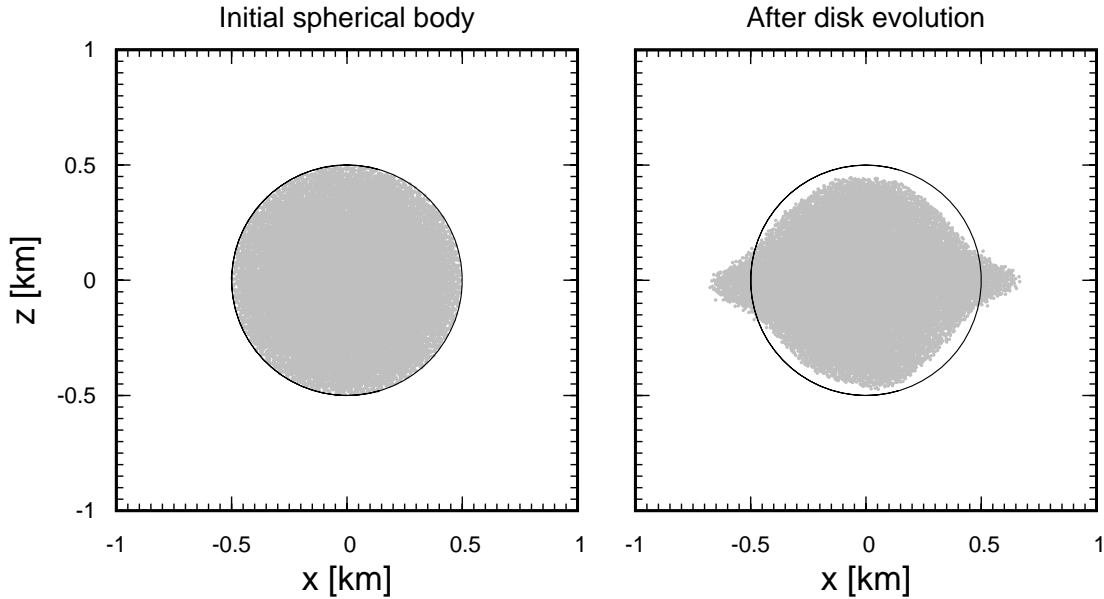}
\caption{Comparison of asteroid shapes between the initial figure ($t=0$h; left panel) and after the disk evolution ($t \sim 700$h; right panel). A black circle indicates a circle with a radius of 500 meters.  \label{fig_ridge}}
\end{figure*}

Our numerical simulations show another interesting feature as a result of disk evolution around a top-shaped asteroid. Because the disk spreads both inward and outward to conserve the angular momentum, the disk particles initially at the inner region re-accrete onto the top-shaped primary. This has led to the formation of an equatorial ridge. 

Other potential physical mechanisms to form an equatorial ridge are discussed in other literature. These are, for example, rotational reshaping and mass movement \citep[e.g.,][]{Wal12,Hir15,Hir20}, re-accumulation of debris after a catastrophic disruption of a parent body \citep{Mic20}, or selective accumulation of ejecta of small impacts at rapidly rotating asteroid \citep{Ike21}.

Figure \ref{fig_ridge} shows an edge on views of the central rubble-pile body at $t=0$h (left panel) and at the end of our simulation without disk particles and formed moons ($t \sim 700$h; right panel). A significant shape change can be seen from the initial spherical shape to a top-shaped figure with a prominent axisymmetric equatorial ridge. This is a direct consequence of the re-accretion of equatorial disk particles onto the top-shaped primary. 

Many of the top-shaped asteroids and relatively spheroidal bodies in the near-Earth region are reported to have an equatorial ridge \citep{Ben15}. JAXA's Hayabusa2 mission revealed that asteroid Ryugu is a top-shaped asteroid with an equatorial ridge \citep{Wat19}. NASA's OSIRIS-REx mission also reported that asteroid Bennu has an equatorial ridge \citep{Wal19}. Another top-shaped asteroid, Didymos, has also an equatorial ridge as well as a moon, Dimorphos \citep{Nai20}.

We importantly note that the axis ratio of $c/a$ and the equatorial ridge seen in Fig.~\ref{fig_ridge} are smaller and much more prominent than the observations of, for example, Ryugu, Bennu, and Didymos \citep{Wat19,Lau19,Nai20}. Our SPH simulations would be limited in their ability to precisely demonstrate the detailed accumulation processes of particles in the equatorial ridge; because our SPH simulations assumed and employed, for example, the same size and simple physical properties among all particles. We also emphasize the importance of studying the long-term geological evolution (e.g., including degradation) of the ridge. Continuous micrometeoroid impacts and/or thermal fatigue would also change the ridge shape. Therefore, further studies of the detailed ridge formation process as well as the post-formation geological processes need to be done to fully validate our ridge formation scenario via the landslides followed by the disk evolution. We leave these points to later work.

\subsection{On the diversity of top-shaped asteroids}  \label{sec_diversity}

Some top-shaped asteroids, e.g., Didymos, are found to have a companion moon(s), while others do not (e.g., Ryugu and Bennu). In the discussion below, we focus on asteroids that have (1) top-shaped figures and (2) equatorial ridges, as background considerations. We, then, additionally consider the existence and non-existence of a companion moon around the top-shaped primary. 

Our results together with those of \cite{Sug21} indicate that small rubble-pile asteroids may inevitably experience a landslide due to spin-up by the YORP effect or by other physical processes (e.g., re-accumulation, small impact, and/or a close encounter with a planet). Depending on the acceleration rate of spin-up as well as the effective friction angle, $\phi_{\rm fri}$, of constituent particles of an asteroid, the resultant shape of asteroids via a landslide would change (e.g., a lemon-shape or a top-shape; see Fig.1 of \citealt{Sug21}). 

When the top-shaped primary was formed with the fast spin-up rate and $\phi_{\rm fri} \gtrsim 70$ deg, \cite{Sug21} numerically demonstrated that about $\sim 10$wt.\% of the primary is generally ejected. When the initial disk mass is $\sim 10$wt.\% or larger, the mass of the moon would be linearly scaled with the disk mass \citep{Ida97,Kok00}, and thus a change in the resultant disk mass within the same order of magnitude does not significantly change the mass of the moon (would be on the order of $\sim 1$wt.\% of the primary as observed for the top-shaped binary asteroids). In contrast, when the effective friction angle is smaller than the above value, the shape deformation became more gradual, and a lemon-shaped primary was formed. In this case, the ejected mass tends to be smaller (e.g., $\sim 3$wt.\% and $\sim 5$wt.\% for $\phi_{\rm fri}=40$ deg and $\phi_{\rm fri}=60$ deg, respectively), although more detailed studies on throughout parameters may be needed \citep[see Fig.~1 of][]{Sug21}. In these cases of smaller initial disk masses, the mass of the resultant moon depends more strongly on the disk mass and much smaller moons can be produced from the disk \citep{Hyo15}.

Importantly, as a natural consequence of a landslide, a particle disk is formed around the primary. Once a particle disk exists, the disk inevitably spreads and forms moon(s) as long as the disk is massive enough. The final mass and orbital configurations of the moon systems formed through the disk spreading depend on the initial disk mass and tidal parameters. Such dependencies are studied using 1D fluid simulations in our companion paper in the context of the Didymos-Dimorphos system formation (Madeira et al. submitted). In short, the direct consequence of the landslide would be the system of a moon(s) around a central asteroid -- regardless of the top-shaped or the lemon-shaped central primary -- with an equatorial ridge on the primary. 

Then, the question now is can we remove the formed moon if the top-shaped figures of Ryugu and/or Benuu are formed via landslides? This is because these top-shaped asteroids today do not have a moon around them. Ejection via tidal evolution would not be promising because its efficiency significantly decreases as the distance between the primary and the moon becomes larger. 

A potential dominant dynamical mechanism on the binary system may be the binary YORP (BYORP) effect \citep[e.g.,][]{Cuk05,Cuk10,McM10,Jac11}. Although the timescale of orbital separation to eject a companion moon via the BYORP effect can be as small as $\sim 10^5$ years \citep[see Sec.4.3 in \citealt{Sug21} and ][]{Cuk05}, its timescale and direction of the orbital separation (shrinking or expanding) strongly depend on the details of the surface properties of the asteroids and moons \citep{Cuk05,McM10}.

We envision that such a complex dependence on the shapes and surface properties can potentially lead to a diversity of the top-shaped asteroids with and without a companion moon. The moons can be ejected by the BYORP effect in some cases, while the dynamical evolution of the orbital separation may not be efficient in other cases, remaining the system as binary asteroids (Figure \ref{fig_summary}).\\

\section{Summary}\label{sec_summary}
In this study, using the SPH simulations, we studied a continuous dynamical sequence of events from the surface landslide that forms a top-shaped asteroid and a particulate disk to disk evolution (Figure \ref{fig_summary}). We numerically demonstrated that the particle disk formed by a surface landslide would quickly spread and produce moons just outside the Roche limit of the top-shaped primary (within $\sim 300$ hours). The mass of the moon is consistent with what is observed around the top-shaped asteroids (on the order of $\sim 1$wt.\% of the primary). We also demonstrated that an equatorial ridge would be naturally formed because a fraction of the disk particles re-accrete selectively on the equatorial region of the primary. 

Tidal interaction as well as the binary YORP (BYORP) effect between moons and the primary would change the orbital separation between them. The timescale and direction of the orbital separation (shrinking or expanding) strongly depend on the details of the surface properties of the moons and the primary. This indicates that a long-term orbital evolution could produce diverse moon systems around the top-shaped asteroids.

We envision that top-shaped Ryugu and Bennu could once have an ancient moon that was later lost as a result of a successive moon's orbital evolution. Alternatively, other top-shaped asteroids today, such as Didymos, have a companion moon. In these cases, no significant orbital evolution of the moon may have occurred that would result in its loss.

Our study focused on the top-shaped asteroids. However, our results of the consequences of a surface landslide -- the moon formation and the equatorial ridge formation -- can be qualitatively (but not always quantitatively) applied to any small rubble-pile asteroids near the breakup rotation period. Indeed, \cite{Sug21} showed that changing the effective friction angle and/or the spin-up rate results in different patterns of the deformation mode with a variety of disk formation (see their Fig.~1). Once a particle disk is formed, the moon and the equatorial ridge would be naturally and inevitably formed. The long-term evolution, then, would lead to a diverse asteroid system \citep[see also][]{Cuk07,Jac11,Cuk21}.

Further development of the theoretical and modeling research, especially on the BYORP effect, together with a better understanding of the surface and particle properties of individual asteroids would be needed to further constrain and validate the results presented in this study. Further studies on the accumulation and post-formation geological processes of the equatorial ridge are also demanded. Analysis of the return samples by JAXA's Hayabusa2 and NASA's OSIRIS-REx as well as data that would be obtained by NASA's DART and ESA's HERA missions would help us to better understand the nature and evolution of the top-shaped asteroids.

\begin{acknowledgments}
R.H. acknowledges the financial support of MEXT/JSPS KAKENHI (Grant Number JP22K14091). R.H. also acknowledges JAXA's International Top Young program. K.S. acknowledges the financial support of JSPS, Japan KAKENHI Grant (JP20K14536, JP20J01165). Numerical simulations in this work were carried out on the Cray XC50 supercomputer at the Center for Computational Astrophysics, National Astronomical Observatory of Japan.

\end{acknowledgments}


\bibliography{citation}{}
\bibliographystyle{aasjournal}



\end{document}